\documentclass[%
 reprint,
 amsmath,amssymb,
 aip,
]{revtex4-2}

\usepackage{graphicx}
\usepackage{dcolumn}
\usepackage{bm}
\usepackage[version=4]{mhchem}
\usepackage{siunitx}

\begin{document}

\preprint{APS/123-QED}

\title{Ferroelastic control of magnetic domain structure: direct imaging by Magnetic Force Microscopy}

\author{S. D. Seddon}
\affiliation{Institut für Angewandte Physik, TU Dresden, D-01069 Dresden, Germany}
\author{C. R. S. Haines}%
\affiliation{%
 Department of Mineral Science, University of Cambridge, Downing Street, Cambridge, CB2 3EQ, United Kingdom
}

\author{T. P. A. Hase}
 \affiliation{%
 Department of Physics, University of Warwick, Coventry, CV4 7AL, United Kingdom
}

\author{M. R. Lees}
 \affiliation{%
 Department of Physics, University of Warwick, Coventry, CV4 7AL, United Kingdom
}%
\author{L. M. Eng}
\affiliation{Institut für Angewandte Physik, TU Dresden, D-01069 Dresden, Germany}
\affiliation{ct.qmat: Dresden-Würzburg Cluster of Excellence—EXC 2147, TU Dresden, 01062 Dresden, Germany}

\author{M. Alexe}
 \affiliation{%
 Department of Physics, University of Warwick, Coventry, CV4 7AL, United Kingdom
}

\author{M. A. Carpenter}
\affiliation{%
 Department of Mineral Science, University of Cambridge, Downing Street, Cambridge, CB2 3EQ, United Kingdom
}%

\date{\today}

\begin{abstract}
Pyrrhotite, Fe$_7$S$_8$, provides  an example of exceptionally strong magnetoelastic coupling through pinning of ferromagnetic domains by ferroelastic twins. Using direct imaging of both magnetic and ferroelastic domains by magnetic force microscopy (MFM), the mechanism by which this coupling controls local magnetic switching behaviour of regions on the pyrrhotite surface is revealed, and leads to quantitative fitting of field dependent MFM phase shifts with bulk magnetometry data. It is shown that characteristic inflection points in the magnetometry data along certain direction, in particular $[\overline 120]^*_h$ of the hexagonal parent structure, are in fact caused by ferroelastic pinning of the magnetic moments.
\end{abstract}

\maketitle

The mineral pyrrhotite, Fe$_7$S$_8$, has been of interest to the paleomagnetic research community on account of the fact that it has a large remanent magnetisation \cite{direen_strong_2008}. Naturally occurring materials that exhibit this property can be crucial for understanding the  magnetic and tectonic history of the Earth\cite{hall_pyrite-pyrrhotine_1986}. Beyond terrestrial paleomagnetic applications, pyrrhotite is also of interest due to its presence in meteorites which originated from Mars \cite{bradley_analysis_1988,zolensky_iron_1995, rochette_pyrrhotite_2001}. Given the violent entry to earth, where large pressure and heat changes are inevitable, specifically understanding the role that strain can have on the relevant magnetic properties is essential to interpreting meteorite samples \cite{kontny_mineralogy_2000,gattacceca_effects_2007,louzada_effect_2007,gilder_anatomy_2011}. Pyrrhotite has also recently been identified as a contender for improving the efficiency of Li-ion batteries \cite{guo_porous_2020}. In the context of the present study, it provides a remarkable example of how the magnetic properties of an effectively multiferroic (ferroelastic/ferrimagnetic) material can be determined by the configuration of ferroelastic domains.

\begin{figure*}[htb!]
	\centering
	\includegraphics[width=\textwidth]{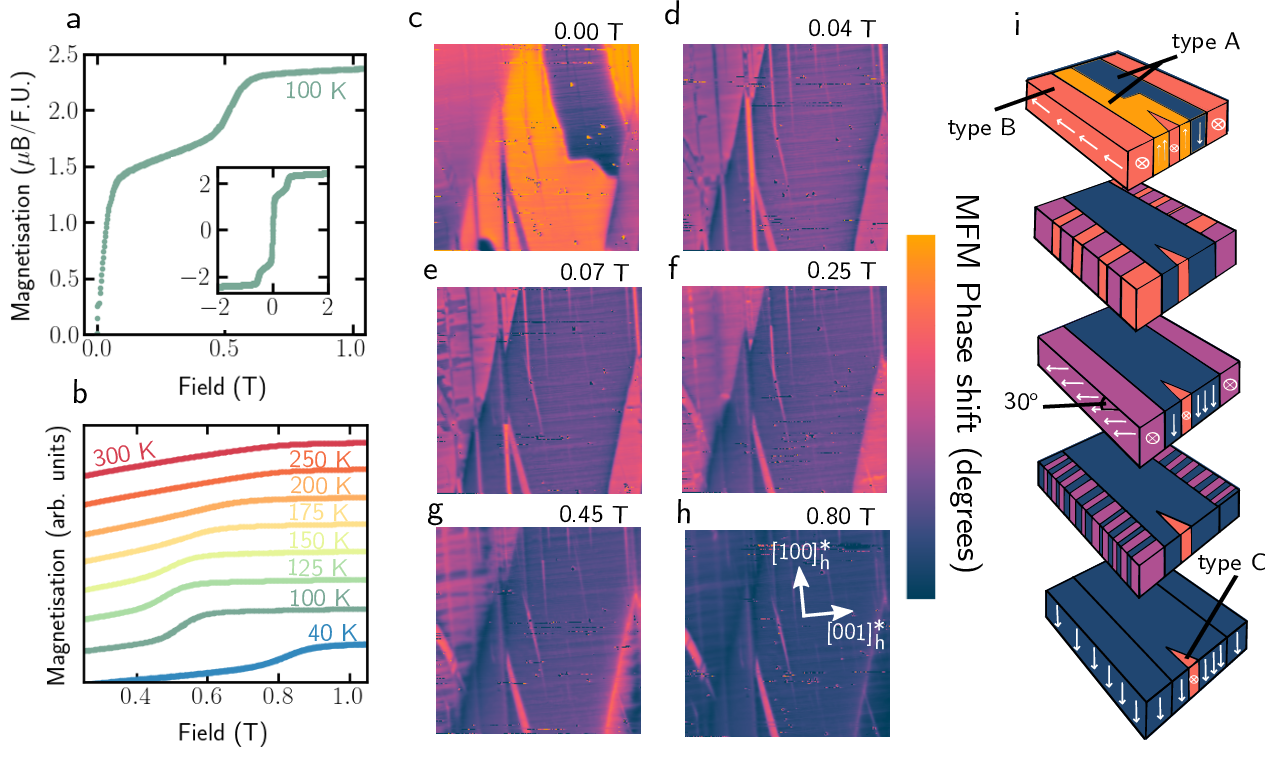}
	\caption{\textbf{a} Magnetisation data of \ce{Fe7S8} acquired at \qty{100}{\K} with field applied parallel to $[\overline{1}20]^*_h$. There is a sub-saturation inflection beginning at \qty{0.5}{\tesla} before saturation is approached at \qty{0.7}{\tesla}. The full magnetisation loop is shown in the inset. \textbf{b} Temperature dependence showing how the inflection point in \textbf{a} becomes most apparent below \qty{175}{\K}. \textbf{c-h} Field dependent MFM images of a surface of the pyrrhotite single crystal cut perpendicular to $[\overline{1}20]^*_h$. Successive images were obtained with increasing positive field,  applied normal to the surface after a negative \qty{4}{\tesla} saturating field had been applied. \textbf{i} Schematic interpretation of the sequence of images with increasing field shown in \textbf{c-h}, including the proposed orientation of moments within the surface layer.}
	\label{fig:MFMdata}
\end{figure*}
There is an intimate relationship between atomic structure and magnetic structure in each of the different superstructure forms of natural pyrrhotites. \citet{powell_structure_2004} provides seminal results of structural refinements of the most common, 4C form, so-called because it has a superlattice repeat which is four times that of the c-dimension of the parent NiAs structure due to ordering of vacancies on the cation site. Vacancy ordering and magnetic ordering in the 4C structure combine to give potentially complex domain structures with contrasting dynamical properties. The change in point group symmetry due to vacancy ordering, $6/mmm \rightarrow  2/m$, occurs by the development of two symmetry-breaking shear strains to produce ferroelastic twin domains with up to six different orientations. If there are six possible domain orientations for vacancy ordering there must also be six possible orientations for the ferrimagnetic moment \cite{haines_magnetoelastic_2020}, each of which can have positive and negative directions. It follows that microstructures associated with the symmetry change $P63/mmc \rightarrow C2'/c'$ must contain both purely magnetic \ang{180} domain walls, and a diversity of walls which are both magnetic and ferroelastic. 

Individual ferroelastic twins have been observed on a scale of $\sim$ \qtyrange{1}{20}{\mm} in natural crystals by reflected light microscopy (e.g. \citet{bennett}) and electron back scattered diffraction\cite{haines_magnetoelastic_2020}, and at a sub-optical scale by transmission electron microscopy \cite{putnis_observations_1975, posfai_pyrrhotite_2000, jin_atomic-scale_2021}. Although this twinning is nominally ferroelastic, it is unlikely that classical switching in response to external fields will be observed at ambient conditions because any change of orientation requires diffusion of vacancies and is expected to be negligibly slow at temperatures below $\sim$~\qty{400}{\K}\cite{haines_magnetoelastic_2020-1}. O'Reilly et al.\cite{oreilly_magnetic_2000} reported that lamellar magnetic domains observed in polished surfaces of polycrystalline samples using the Bitter pattern technique, have a strong tendency to be aligned normal to the crystallographic c-axis due to the strong magnetocrystalline anisotropy of the structure. They found, also, that similar \ang{180} domain structures in other pyrrhotite samples were highly mobile under the influence of an external magnetic field.

While probing the crystallographic direction and temperature dependence of the magnetic hysteresis of 4C pyrrhotite, \citet{haines_magnetoelastic_2020} observed that, for magnetic fields applied along the $[\overline{1}20]^*_h$ direction of the parent hexagonal structure, the magnetisation did not immediately reach saturation, but had several energy barriers to overcome. Such step-like behaviour of the magnetisation in response to an applied field is not an entirely unusual occurrence in the magnetism community, although the underlying reasons are vastly varied. For example \ce{Ca3CoO6}, an Ising spin chain system,  exhibits several smooth steps on its way to complete saturation in a similar appearance to pyrrhotite. The mechanism remains in discussion, with some groups proposing a quantum tunnelling of the magnetisation \cite{maignan_quantum_2004} and others proposing a kind of ferrimagnetic switching where different magnetic centres on the spin chain unit cell switch at different fields \cite{kudasov_steplike_2006} - both potential explanations attributed to individual spin behaviors, and not moments acting as a larger assembly. The unusual magnetic switching behaviour in pyrrhotite has also generated some discussion, regarding specifically apparent differences in behaviour of the net moment  \cite{fillion_neutron_1992,koulialias_torque_2018}. This difference in saturated moment, reported to be an effective difference in final rotation of between \ang{14} and \ang{29}, has been ascribed to a ``hidden anisotropy generated by defects" \cite{fillion_neutron_1992,koulialias_torque_2018,kind_domain-wall_2013,bin_magnetic_1963}. It has been recognised that twinning must influence the magnetic properties\cite{volk_changes_2018}, but the ferroelastic nature of 4C pyrrhotite has not considered directly, and the proposed defects remain unidentified, motivating further investigation.

The 4C pyrrhotite sample used for magnetic measurements was part of the same single crystal as pieces used in previous work\cite{haines_magnetoelastic_2020, haines_magnetoelastic_2020-1}. The crystal came from the mineral collection of the South Australia Museum and had originally been collected from a mine in Mexico. For convenience, crystallographic faces, their surface normal and other directions in the crystal are described by referring to the hexagonal setting of the parent structure. In this context, $[100]^*_h$, $[\overline{1}20]^*_h$ and $[001]^*_h$ of the hexagonal structure are equivalent to $[100]_m$, $[010]_m$ and $[001]_m$ of one of the possible twin orientations of the 4C monoclinic structure. A surface perpendicular to $[\overline{1}20]^*_h$ was prepared by polishing with an Allied Multiprep system to nm surface flatness. The direction of the surface normal was confirmed to be  $[\overline{1}20]^*_h$ by X-Ray Diffraction performed on a Panalytical X'Pert Pro MRD. The plane of this surface contains the orthogonal directions $[001]^*_h$ and $[100]^*_h$. Magnetometry data were acquired using an Oxford Instruments Vibrating-Sample Magnetometer (VSM).

Magnetic Force Microscopy images were acquired in a dual-pass mode. In this approach, the first pass is a normal non-contact mode in order to attain topographic contributions, which are then used to remove topographic interference on the second pass, a ~100~nm lifted pass of the same line scan, now detecting only the magnetic contributions. Temperature and field were varied using a low temperature AFM (Attocube attoLiquid 2000) provided with interferometric SPM system and \qty{9}{\tesla} axial / \qtyproduct{1x1}{\tesla}  vector magnetic field capability and \qtyrange{1.8}{300}{\K} temperature range. Nanosenors PPP-MFMR cobalt coated tips, with ~80 kHz resonance, were used. Python was used for image processing, data analysis and plotting. Resonance amplitude and frequency were adjusted to ensure consistency between scans at different applied magnetic fields.

Unusual plateau like features had previously been identified in magnetometry data for \ce{Fe7S8} 4C samples when the magnetic field was applied along $[\overline{1}20]^*_h$, ie normal to the surface prepared for MFM measurements. A typical example of the plateau at 100 K can be seen in Fig.~1a. The complete loop is shown in the inset, displaying a negligible hysteresis. Instead of following a typical `S'-shaped loop, there is a sub-saturation kink: initial linear progression upwards of \qty{0.1}{\tesla} is followed by a second inflection at \qty{0.5}{\tesla} and trend towards an apparent saturation before finally  flattening. This second inflection point is not present in magnetisation curves collected at room temperature but occurs at increasing applied magnetic field in curves collected at progressively lower temperatures below \qty{200}{\kelvin} (Fig.~1b). Below the \qty{40}{\kelvin} Besnus transition point the pattern of magnetisation changed significantly to include opening of the hysteresis loop (see Figure~4c of \citet{haines_magnetoelastic_2020-1}). As described below, stabilisation of the inflection point is likely due to the increasing dominance of ferroelastic pinning in the magnetic behaviour. On this basis, 100 K was chosen as an appropriate temperature at which to focus the magnetic imaging investigations.
\begin{figure}[tb!]
	\centering
	\includegraphics[width=\linewidth]{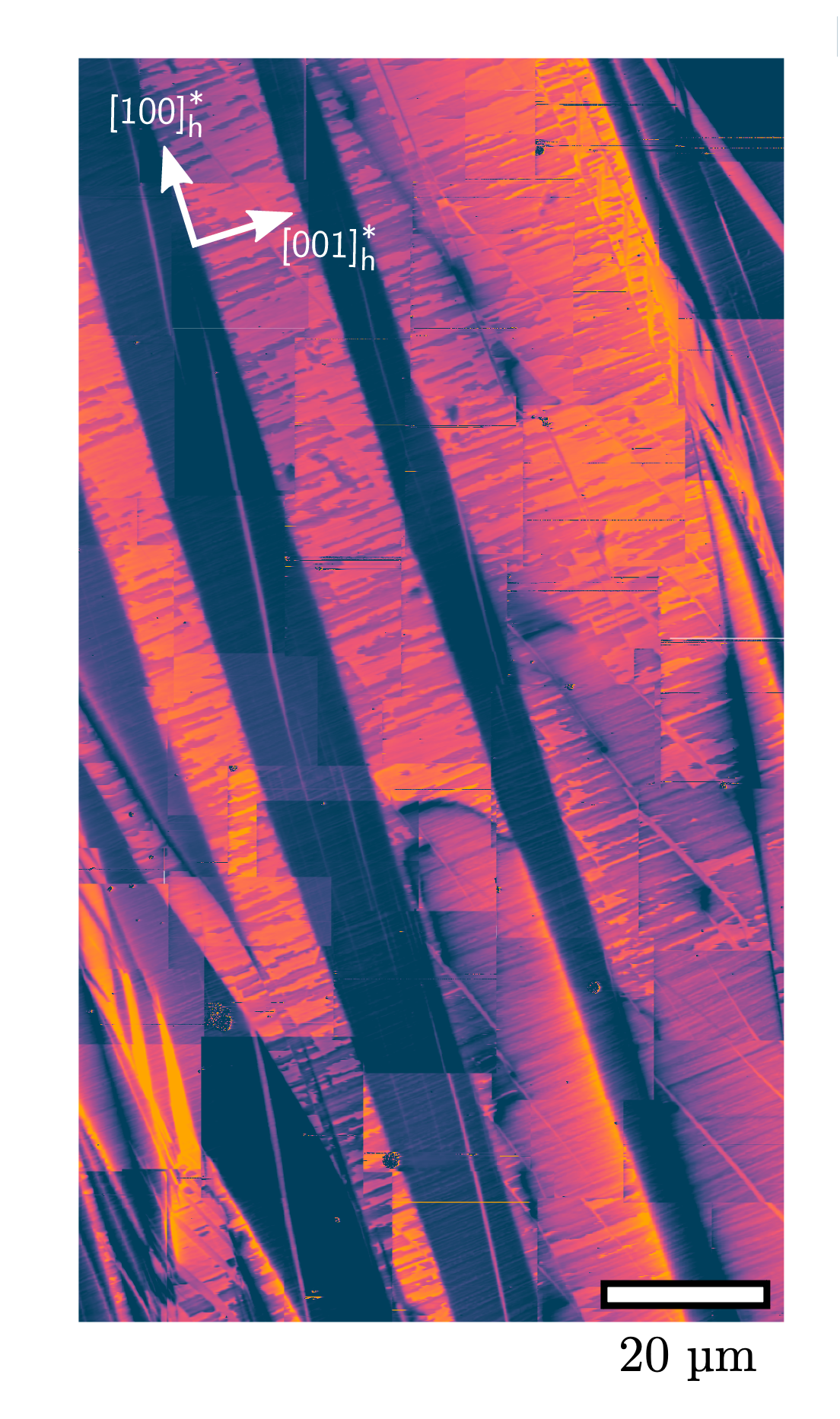}
    	\caption{Stiched MFM scans of \ce{Fe7S8} displaying the distribution of the three types of domains revealed by applying a small magnetic out-of-plane field of 0.07~T. A clear preferential strip direction of the $[100]^*$ (right to left diagonal), as expected}
	\label{fig:MFM_map}
\end{figure}
Prior to acquisition of MFM images, the sample was cycled in magnetic field to remove any virgin magnetisation effects and then magnetised downwards in a \qty{-4}{\tesla} field. Images acquired with increasing positive field from \qtyrange{0}{8}{\tesla} revealed three distinct switching regimes. Exemplar images can be seen in Fig.~1c-h with a colour scale to represent the MFM phase shift in each of the many magnetic domains. The images can be described broadly in terms of three ferroelastic domains with domain walls aligned approximately from bottom left towards top right. The central domain will be referred to below as ``type-A'' and the two on each side as ``type-B''. Phase shifts in the two type-B domains, at the left and right hand sides of the image, are denoted by a light pink colour in zero field (Fig.~1c), representing regions with in-plane magnetic moments. Two regions of the central (type-A) domain show a stark contrast of yellow and deep purple, representing the maximal phase shifts measured by MFM. They are interpreted as corresponding to magnetic domains with moments lying along the positive (yellow) and negative (purple) surface normal. Application of a field as small as \qty{0.04}{\tesla} (Fig.~1d) was sufficient to reorient moments in the yellow region
so that the phase shift colour became the same as in the deep purple region, indicating that the moments had all become aligned in the same direction. Application of the \qty{0.04}{\tesla} field did not affect the type-B domains which remained pink (Fig.~1d)

Switching of some magnetic domains to align with the applied field, as seen in the progression from Figure 1c to 1d, is in good agreement with the initial rise in magnetisation below \qty{0.1}{\tesla} seen in the magnetometry data (Fig. 1a). In contrast with this low-field domain switching behaviour, the type-B ferroelastic domains displayed an intermediate phase shift at both zero field and \qty{0.04}{\tesla}. Small downward sub-domains began to nucleate in these after the initial switching of the type-A domain was complete (Fig. 1e, \qty{0.07}{\tesla}). The individual sub-domains grew under the influence of increasing field until a new steady state was reached at \qty{0.25}{\tesla} (Fig.~1f). At this stage, corresponding to a point within the intermediate plateau between \qtyrange{0.07}{0.45}{\tesla} of the magnetic hysteresis curve, the subdomains had alternating pink/purple colours and boundaries lying approximately east-west in the image (corresponding to the $[010]^*_h$ direction).

\begin{figure*}[tb!]
	\centering
	\includegraphics[width=\linewidth]{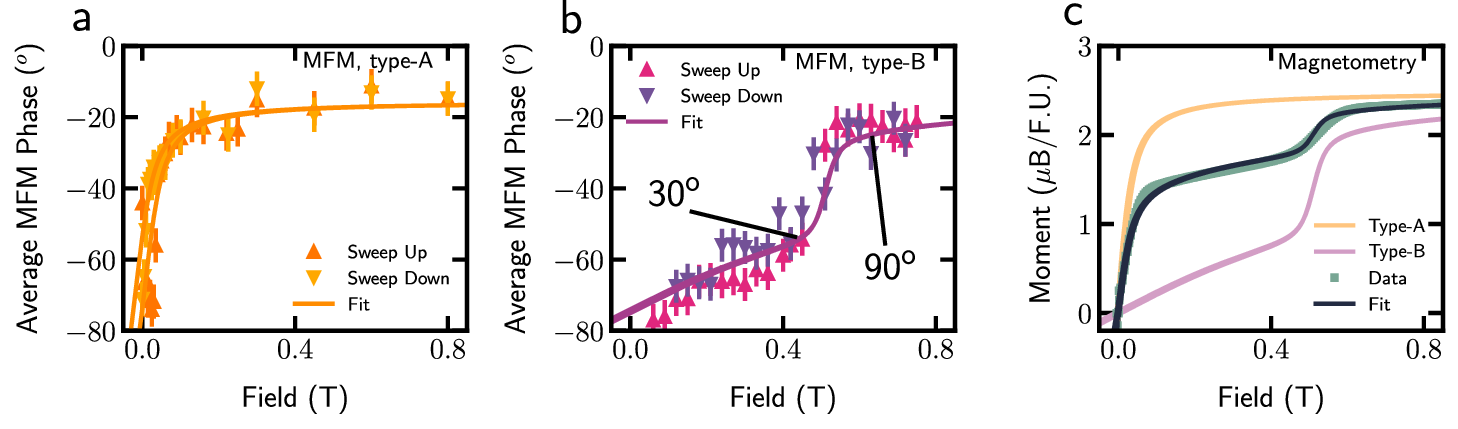}
	\caption{Average phase in \textbf{a} and \textbf{b} are of acquired MFM images for regions selected on the \ce{Fe7S8} sample surface as being entirely type-A and -B respectively. Each data point represents the averaged phase shift from one MFM image, with upward and downward facing triangles data points for upward and downward field sweeps, respectively. Data were fitted simultaneously with a linear combination of both fits (described in text) alongside magnetometry data fitted in \textbf{c}. The final, fitted linear combination of type-A and -B domains (in a $3:2$ ratio) is shown in dark green overlaying the measured magnetometry data (light green). Hysteresis loops for uniform domains of either type-A (yellow) or type-B (purple) are shown for reference.}
	\label{fig:MFMfits}
\end{figure*}
Figure 2 shows the configuration of ferroelastic and magnetic domains, recorded by MFM in \qty{0.07}{\tesla} in more detail on a different part of the surface. Also shown are crystallographic directions within the surface. The predominant orientation of boundaries between the type-A and type-B ferroelastic domains is normal to $[100]^*_h$, suggesting that the domain walls may be parallel to (001). (001) ferroelastic domain walls are also magnetic domain walls. Segments of these walls in some parts of the other A/B boundaries and at needle twins have different orientations, as would be expected for other ferroelastic twin walls cutting the surface obliquely. The walls between alternating pink and yellow/orange in the type-B domains tend to be oriented parallel to $[100]^*_h$, an orientation which implies that they are ferroelastic as well as being magnetic. They are also linear, consistent with being constrained by a ferroelastic shear strain.

Thin, needle-shaped domains aligned approximately north-south in all the images are ``type-C" ferroelastic domains. They did not move in response to the applied field and their pink, intermediate colour in all images throughout the full sequence indicates that their moments remained aligned within the plane of the surface of the sample. Given the high longitudinal magnetic anisotropy expected of a needle shaped ferroelastic domain, it is not unexpected that the magnetic moment points into the tip of each needle, i.e. along its length in this section. Application of fields up to 8 T did not modify these domains in any way.

Further increases in the magnitude of the field led to the type-B domains breaking up into long, striated sub-domains at 0.45 T (Fig.~1g). This corresponds to the beginning of the final step up to saturation in the magnetometry data. The phase shift colour of both type-A and type-B ferroelastic domains became a uniform deep purple colour by \qty{0.8}{\tesla} (Fig.~1h), implying that moments, apart from those in the type-C domains, had finally became aligned in a common direction either exactly parallel to or nearly parallel to the field.

A schematic interpretation of the observed pattern of domain evolution with increasing field is given in Fig.~1i. Type-A domains are shown as initially having moments normal to the surface, while type-B and type-C domains have moments aligned in-plane. In reality the steepest angle for moments to be inclined away the $[100]^*_h$-$[001]^*_h$ surface is \ang{60} but the essential point is that the boundary between the yellow and purple regions in type-A domains is a \ang{180} magnetic domain. The earliest stage of increasing moment with increasing field involves elimination of these. The second stage involves a development of ferroelastic/magnetic domains within the type-B domains such that half the domains have moments lying in-plane and half have moments aligned at an angle of \ang{30} to the surface. In other words, rotation of the moments towards the alignment of the field in the type-B domains does not occur by a continuous rotation. The intermediate plateau of moment vs field is thus a metastable state with a particular configuration of new ferroelastic/magnetic domain walls. Increasing the field further eventually causes all the moments to align with the field and accounts for the second steep increase in moment towards saturation. The in-plane magnetic structure of type-C, needle domain remained unchanged even up to \qty{8}{\tesla}.

The plateau in the evolution of magnetic moment with increasing positive field shown in Fig.~1a matches the pattern of evolution of domains in the images. The plateau also appears in each of the four quadrants of the full hysteresis cycle of increasing, decreasing and changing direction of the field (insert in Fig.~1a), indicating that, overall, the domains evolved in a manner that was fully reversible. None of the primary ferroelastic domain walls moved, leaving the underlying ferroelastic domain structure as a template which constrained how the magnetic domains returned when the field was removed. It follows that the precise pattern of magnetisation for any given crystal must depend substantially on the configuration of ferroelastic domain walls that developed during crystallisation or during cooling from high temperatures. In this context, failure of type-C domains to saturate, due to the high shape anisotropy of the needles, is likely to have been the cause of the different reports of magnetometry derived saturation magnetisation values in the literature \cite{fillion_neutron_1992,koulialias_torque_2018,kind_domain-wall_2013,bin_magnetic_1963}.

In order to quantify the behaviour of the three types of domains, an additional series of images was acquired. By mapping a relatively large area of around \qty{10000}{\micro\meter\squared}, a clear prevalence for stripes of alternating type-A and type-B domains with boundaries generally aligned along $[100]^*_h$ was confirmed. Magnetic domains in this orientation have been identified previously by use of the Bitter pattern technique \cite{oreilly_magnetic_2000, haines_magnetoelastic_2020-1}, but the microstructure reported on here is generally well below the resolution of ferrofluids. At \qty{0.07}{\tesla} all of the type-A domains had saturated and so appeared as a (dark purple) low phase shift. The unsaturated type-B domains, with a lower phase shift in the images, had begun their own, stepped rotation. Two large areas were selected as having predominantly either type-A or type-B domains.

MFM data were also acquired in specific areas of type-A and type-B domains as in successive increments of increasing and decreasing field shown, as shown in Fig.~3. In order to ensure each image was directly comparable, the stability of type-C domains was exploited. Every individual image was phase-adjusted to ensure that the type-C phase shift was the same. Each data point in Figs.~3a and 3b then represents the average phase recorded for type-A and type-B domains, respectively, where each point is an MFM image of the same area.

The data in Fig.~3 confirm the qualitative descriptions provided above. Type-A domains saturated with very low fields and followed a traditional `S'-like loop. Conversely, type-B domains started from a low phase shift but required much higher applied fields to approach saturation. Their reasonably linear contribution to the total magnetisation is punctuated by a sharp transition to a stable point where magnetic saturation was finally reached.

In order to directly compare the effect of having two alternative magnetic switching behaviours present in the material, data were fitted with a modified Langevin function (as in \cite{procter_magnetic_2015,thorarinsdottir_amorphous_2021, seddon_characterization_2024}) of the form

\begin{equation}
    M = M_s \left( \frac{1}{\tanh(\frac{H+H_c}{S})} - \frac{1}{\frac{H+H_c}{S}} \right),
\end{equation}
here $M$ is the moment of the phase being fitted, $M_s$ is the saturated moment, $H$ is the applied field, $H_c$ is the coercive field, and $S$ is the `shape'. Fits were performed with a Marquardt-Levenberg algorithm reducing on $\chi^2$. The shape parameter effectively controls the squareness of the loops but, as this fit is phenomenological, no physical meaning may be derived from it. Directly relating the phase shift of an MFM cantilever to the position or rotation of the magnetic moments is analytically challenging. However, by quantifying each image into directly comparable `values', self consistent with each other, a conversion factor can be applied to fits of the MFM data to scale them to be comparable with VSM data. By linearly combining the fits for type-A and type-B domains according to
\begin{equation}
\label{eqn:Fe7S8-lincomp}
    M_{\text{VSM}}(H) = G_{\text{VSM}}(\gamma_{A} \phi_{A}(H) + \gamma_{B} \phi_{B}(H)),
\end{equation}
allowing the two MFM data sets to be combined and fit to attain a fit of the fit of the VSM data shown in Figure 3c. In equation~2 $M_{\text{VSM}}$ is the VSM field value at any applied field, $G_{\text{VSM}}$ is an arbitrary scaling factor, $\phi _A(H)$ and $\phi _B(H)$ are the average MFM phase values at any field value for type-A and type-B domains respectively, and $\gamma _A$ and $\gamma_B$ are linear scalars (${\gamma_A + \gamma_B = 1}$).

Equation 2 has no free parameters, only scaling terms. Whilst not leading to a completely perfect fit, it is worth pointing out that the type-A and type-B MFM data sets were taken from very small, specific areas unlike the VSM data set which is a global sample measurement. Even with the obvious challenge of a small area measured for MFM, visually, the fit is good -  reflecting the key features and plateaus in the VSM data. Also plotted on Figure~3c are the type-A and type-B fits with $\gamma_A=0$ and $\gamma_B=1$ or vice versa. This effectively compares the extrema where the sample only consists of one type of ferroelastic/ferromagnetic domain. It was found from the fitting procedure that more type-A than type-B was present in a ratio of $3:2$, i.e. 60\% of the sample was made up of type-A domains. This number does not include the type-C domains whose moment lies entirely in the plane and did not undergo reversal. This is a particularly interesting result when considering the seeming abundance of ``type-B domains'' in Figure~2. Naturally, scanning only a specific area of the sample is not going to yield bulk results; however, by combining local results simultaneously with alternative bulk measurements reveals further information about the local microstructure.

The combination of MFM observations and VSM magnetic data indicate that, although new ferroelastic/magnetic domain walls may be created in some domains when a magnetic field is applied, the original ferroelastic domain walls do not move. In a geological context, palaeomagnetic moments recorded in multidomain 4C crystals should therefore provide a robust source of palaeomagnetic remanence, and needle-shaped domains should be the most robust. However, any process which disrupts the original distribution of ferroelastic domains should then have a dramatic impact on the orientation and magnitude of the bulk magnetic moment. This would be a factor in relation to experimental findings that both shock and static pressure can cause demagnetisation of pyrrhotite, and the proposal that loss of magnetisation of the Martian crust was due to impact of meteorite bodies\cite{louzada_effect_2007, louzada_shock_2010, rochette_impact_2003} . It would be worth examining the distribution of ferroelastic domains in 4C grains before and after such experiments to test whether redistribution of the domain walls by shearing provides a significant contribution to the loss. For example, \citet{mang_shock_2013} observed abundant deformation twins on a scale of \qty{50}{\nano\meter} in crystals that had been subjected to shock up to a pressure of \qty{30}{\giga\pascal}.

Magnetic imaging has allowed direct observations of the interaction of a particular class of ferroelastic domains, i.e. arising from cation/vacancy ordering, with magneticdomains and the response of both to an external magnetic field. The two predominant twins have been labelled as type-A, with an easy axis along or slightly inclined to the surface normal of the crystal, and type-B, with moments free to rotate within in-plane sample directions but with easy axes at \ang{30} increments out of the plane. It is this rotational behaviour that is responsible for the unusual inflections in the hysteresis along this specific field direction. Finally type-C domains do not reach saturation in an out-of-plane field, due to the high structural anisotropy arising from their needle-like shape which pins the magnetic moments into the plane of the sample, along the length of the needle. These are identified as the likely cause of the different saturation magnetisation that is recorded between different natural samples, due to a difference in abundance of this specific type of domain. It is clear that ferroelastic domains in pyrrhotite have a fundamental influence on the magnetic response to an external field, and there is no sign that walls between any of these can be moved by fields up to \qty{8}{\tesla} at \qty{100}{\kelvin}. In a wider context, simultaneous imaging of both ferroelastic and magnetic domains by MFM, together with the particular properties of pyrrhotite provide a window into how magnetic properties of functional materials with multiple instabilities might be engineered through the choice of microstructure. Furthermore, the extrapolation of MFM, a primarily locally resolved technique with the measured bulk properties highlights that with appropriate treatment or large data volumes,  scanning probe microscopy techniques can be used to identify and correlate bulk behaviors. 

\textbf{Acknowledgements}\\
The work was partly supported by the Engineering and Physical Sciences Research Council (EPSRC, UK) through grant nos EP/M022706/1 (M.A.) and EP/T027207/1 (M.A.)

\textbf{Conflicts of interest}\\
The authors have no conflicts of interest to disclose

\textbf{Data availability}\\
The data that support the findings of this study are available
from the corresponding author upon reasonable request.

\textbf{Bibliography}
\bibliography{pyrr_bib}

\end{document}